\documentclass[a4paper,11pt]{article}
\pdfoutput=1 

\usepackage{jinstpub} 

\title{\boldmath Charge sharing in single and double GEMs}

\author[a,b,1]{Promita Roy,\note{Corresponding author.}}
\author[c]{Purba Bhattacharya,}
\author[a,b]{Supratik Mukhopadhyay,}
\author[a,b]{Nayana Majumdar}

\affiliation[a]{Saha Institute Of Nuclear Physics, Kolkata-700064, INDIA.}
\affiliation[b]{Homi Bhabha National Institute, Mumbai -400094, INDIA.}
\affiliation[c]{Departimento Di Fisica, Istituto Nazionale di Fisica Nucleare, Monserrato CA, Italy.}

\emailAdd{promita.roy@saha.ac.in}

\abstract{The Gas Electron Multiplier (GEM) has become a widely used technology for high-rate particle physics experiments like COMPASS, LHCb and are being used as the readout system for the upcoming upgrade version of other experiments such as ALICE TPC. Radiation hardness, ageing resistance and stability against discharges are main criteria for long-term operation of such detectors in high-rate experiments. In particular, discharge is a serious issue as it may cause irreversible damages to the detector as well as the readout electronics. The charge density inside the amplification region is one of the limiting factors for detector stability against discharges. By using multiple devices and thus, sharing the electron multiplication in different stages, maximum sustainable gain can be increased by several orders of magnitude. A common explanation for this is connected to the transverse electron diffusion, widening of the electron cloud and reducing the charge density in the last multiplier. However, this has not been verified yet. In our work, we are using Garfield simulation framework as a tool to extract the information related to the transverse size of the propagating electron cloud and thus, to estimate the charge density in the GEM holes for multiple stages. For a given gas mixture, we will present the initial results of charge sharing using single and double GEM detectors under different electric field configurations and its effect on other measurable detector parameters such as single point position resolution.}


\arxivnumber{1234.56789} 

\proceeding{XV$^{\text{th}}$ Workshop on Resistive Plate Chambers and Related Detectors RPC2020\\
  10-14 February 2020\\
  Rome, Italy.}

\begin{document}
\maketitle
\flushbottom

\section{Introduction}
\label{sec:intro}

Gas Electron Multiplier (GEM) is one of the popular Micro Pattern Gaseous Detectors (MPGDs) which are high granularity gaseous ionization detectors with very small gap between cathode and anode electrodes\cite{a}. High granularity of these detectors offers good position resolution and small gap between the electrodes offers high rate capability, thus,  they are widely used for tracking and timing purposes in various high energy physics experiments.

GEM detectors have very thin dielectric foil(s) coated with copper on both sides (called GEM foil), placed in between cathode and anode. This GEM foil has numerous holes, across which high voltage is applied to allow the multiplication of electrons. Thus, a GEM detector has a separate drift gap, multiplication gap(s), transfer gap(s) and an induction gap. However, the applied high voltage can result in electric discharge across the foil under several circumstances, which in turn can affect the long-term operation of these detectors. It has been proposed that the development of electric discharge across GEM foils depend on the charge density within the configuration\cite{b}. This motivates us to explore the influence of detector geometry and field configuration on charge sharing within GEMs and study the dependence of various other figures of merits on this parameter.

\section{Simulation Tool}
We have used the Garfield\cite{c} framework as simulation tool for the detailed study of gaseous detectors where we have used Monte Carlo and microscopic tracking methods for simulating the charge transport. The 3D electrostatic field has been estimated by neBEM\cite{d}. Apart from these, HEED\cite{e} and MAGBOLTZ\cite{f} have been used for simulating primary ionization and transport properties respectively.

\section{Results and Discussion}
A single GEM detector with 1.5mm drift and 500$\mu$m induction gap has been studied. In addition, a double GEM detector with drift gap of 3mm, transfer gap of 2mm and induction gap of 1mm has been studied. For both the detectors, standard GEM foils have been considered with 50$\mu$m  thick kapton foil, 5$\mu$m copper coating, 140$\mu$m hole pitch and 70$\mu$m hole diameter.

\begin{figure}[htbp]
	\begin{minipage}{6cm}
		\includegraphics[scale=.23]{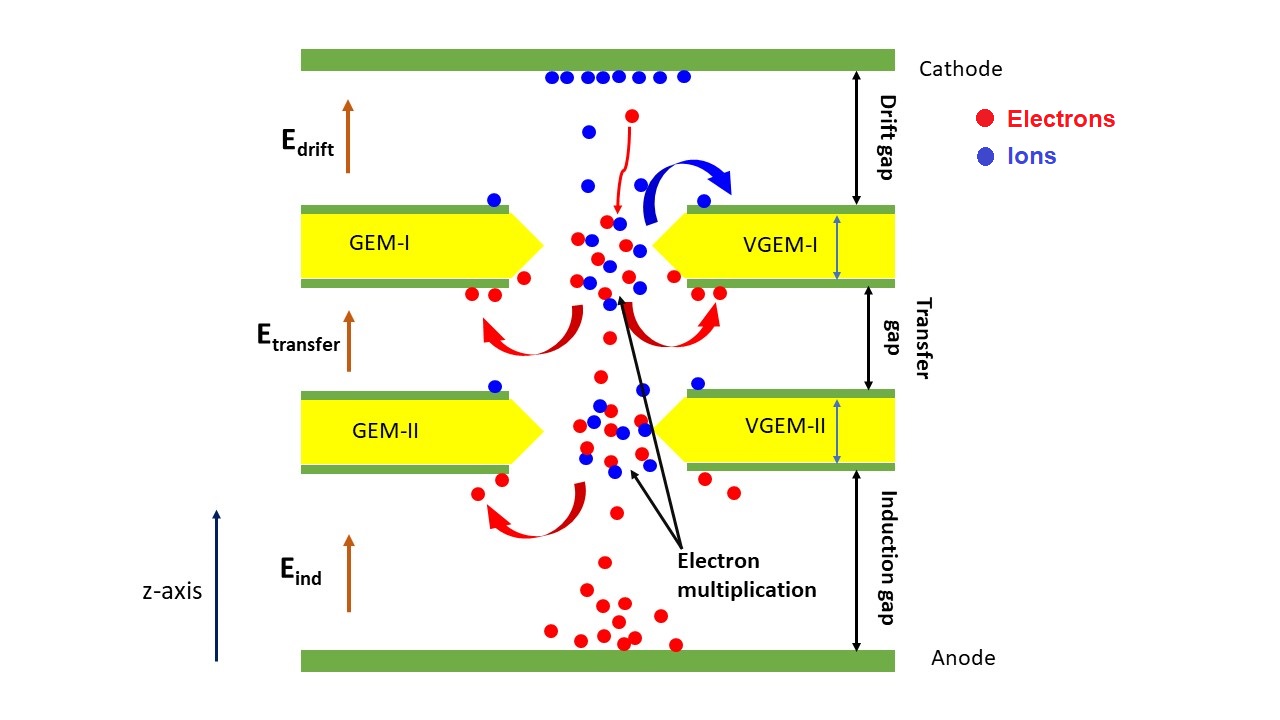}
		\caption{\label{fig1} 2D model of a double GEM.}
	\end{minipage}
	\hspace{1cm}
	\begin{minipage}{6cm}
		\centering
		\includegraphics[scale=.27]{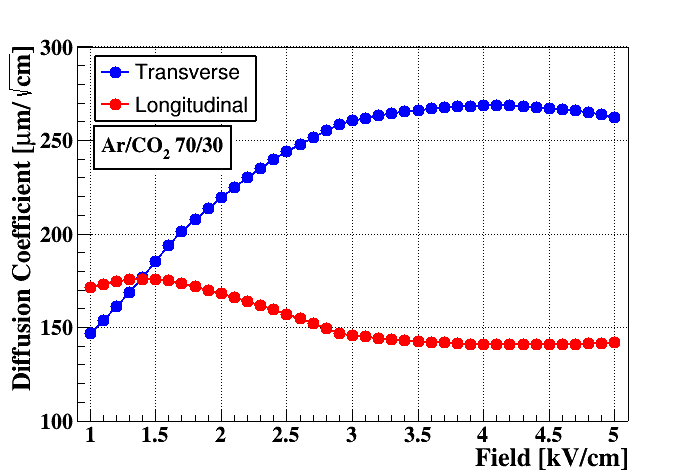}
		\caption{\label{fig2} Variation of diffusion coefficients with electric field}
	\end{minipage}
\end{figure}

We have simulated the spread of electrons, estimated average charge density at different holes and computed single point spatial resolutions for single and double GEM detectors being operated with a mixture of Ar and CO\textsubscript{2} gases in the ratio 70:30. In figure\ref{fig1}, a double GEM detector has been illustrated. Some of the important processes and nomenclatures are also shown in this figure. Variation of transverse and longitudinal diffusion coefficients with electric field has been plotted in figure\ref{fig2}.

\subsection{Spread of electron cloud}
\begin{figure}[htbp]
	\centering
	\includegraphics[scale=.25]{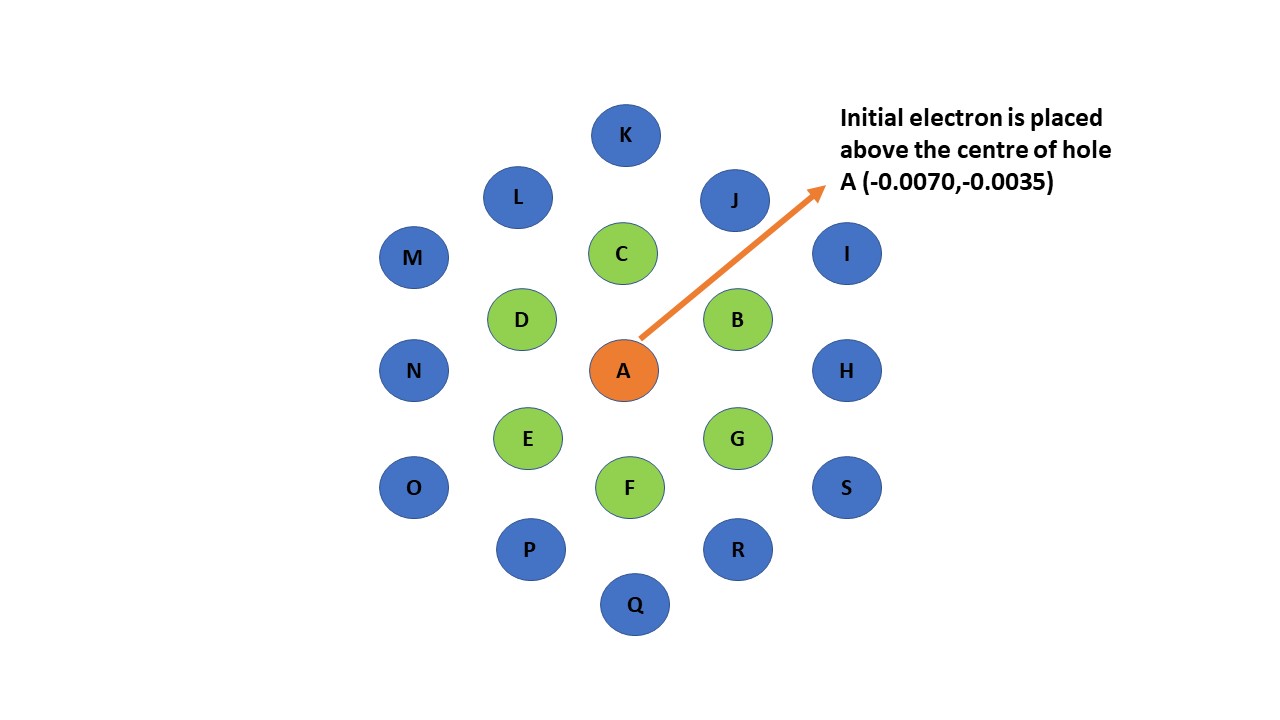}
	\caption{\label{fig3} Schematic of holes of a GEM foil.}
\end{figure}

Starting from a single prefixed location, the spread of the electron cloud has been studied at various locations of a given detector. For a single GEM detector, the spread has been estimated at the top and bottom of the GEM foil and at the anode readout by counting the number of electrons and their positions on these surfaces. For a double GEM detector, the same has been estimated at all the four surfaces of the two GEM foils and on the anode readout. In addition to the surfaces, numbers of electrons in each GEM hole have been computed. For this purpose, the geometry of a GEM foil has been considered as shown in figure\ref{fig3} and an event corresponds to an initial electron starting at 1.5mm just above the centre of hole A.

\begin{figure}[htbp]
	\begin{minipage}{6cm}
		\centering
		\includegraphics[scale=.17]{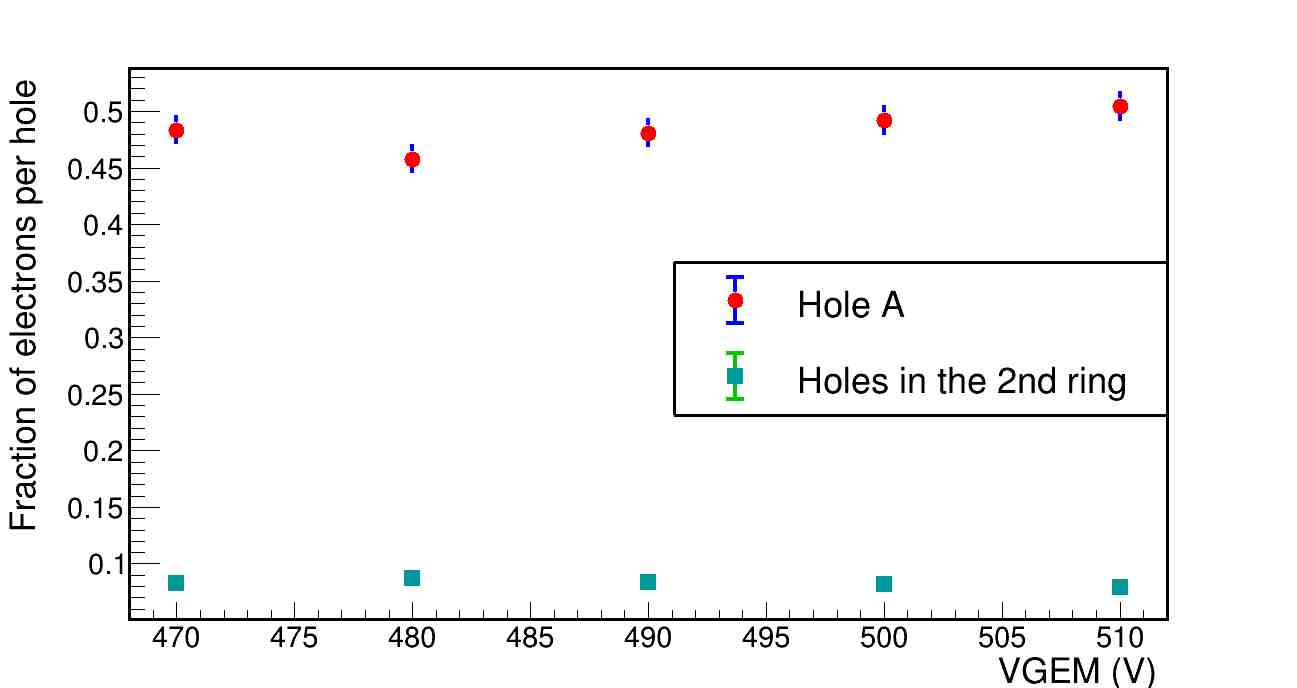}
		\caption{\label{fig4} Fraction of electrons in the central and its adjacent holes of GEM foil in a single GEM.}
	\end{minipage}
	\hspace{1cm}
	\begin{minipage}{6cm}
		\centering
		\includegraphics[scale=.175]{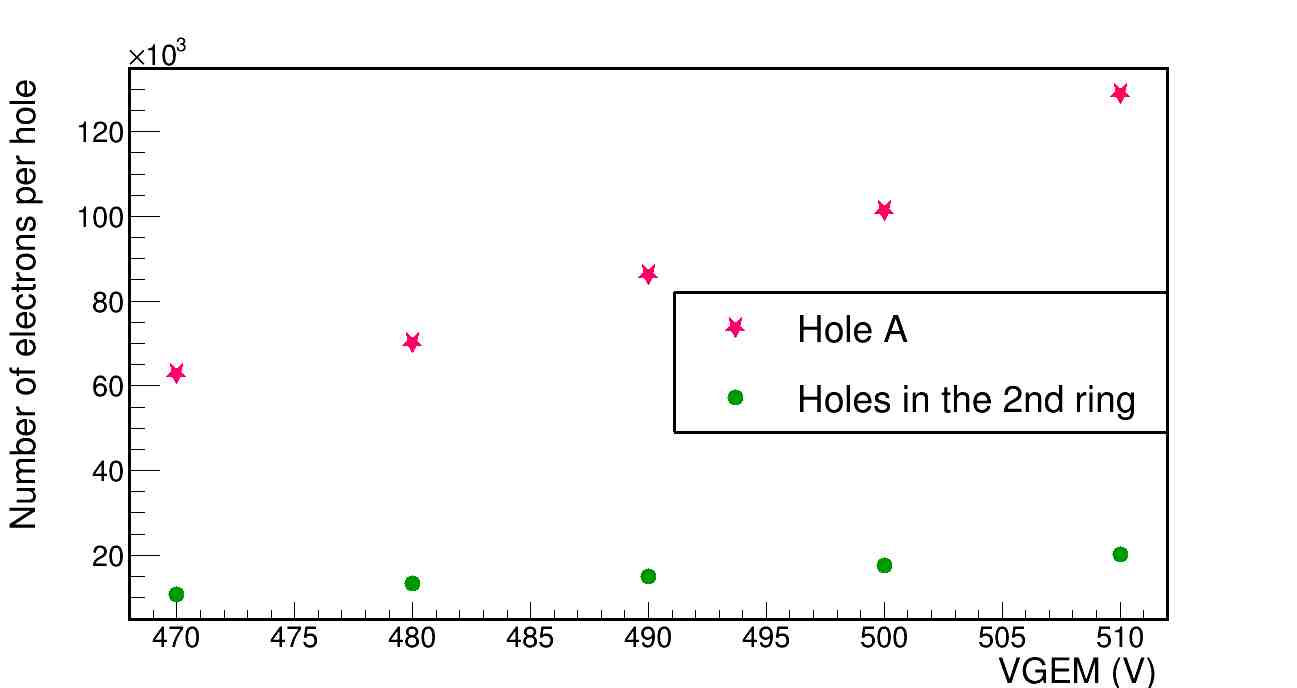}
		\caption{\label{fig5} Charge density in different holes of GEM foil in a single GEM.}
	\end{minipage}
\end{figure}

As expected, a large fraction of the total electronic charge in the GEM foil is present in the central hole A, as shown in figure\ref{fig4}, which in turn shows that a large fraction of initial electrons has entered hole A. Only a small fraction passes through the adjacent holes in the second ring. Interestingly, this fraction for the central hole increases slightly with the increase in GEM voltage, possibly due to improved focusing of the electric field and increasing electron multiplication factor($\alpha$). For instance, around 50$\%$ of total electrons pass through hole A when the GEM voltage is 500V, while the number is 45$\%$ when the voltage is 480V. However, once an electron enters a particular hole, the number of electrons in that hole is similar to that obtained at other holes and the number increases as the applied voltage is increased. This is natural, since the voltage across each hole is the same. This fact is presented in figure\ref{fig5}. So, summed over a large number of such events, the electron load on the central hole remains almost ten times the adjacent holes, for single GEM detector.


\begin{figure}[htbp]
	\begin{minipage}{6.5cm}
		\centering
		\includegraphics[scale=.18]{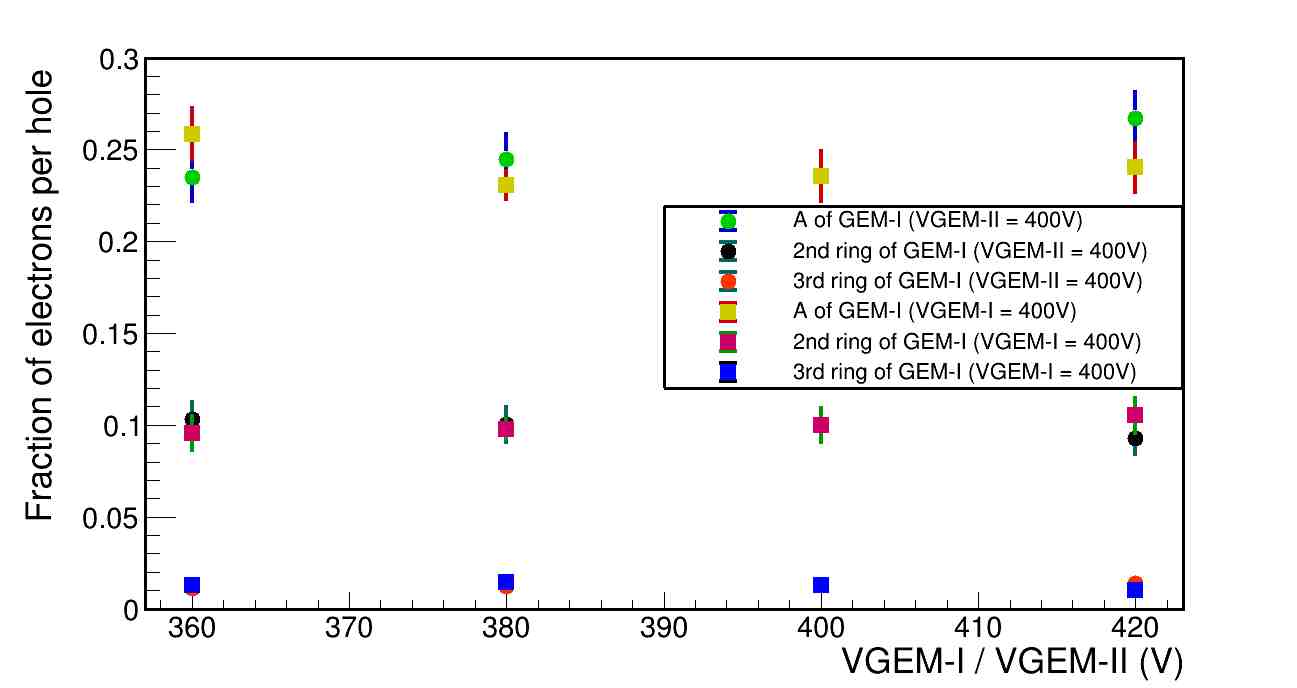}
		\caption{\label{fig6} Fraction of electrons in the central and its adjacent holes in GEM-I for different combinations of VGEM-I and VGEM-II in a double GEM.}
	\end{minipage}
	\hspace{1cm}
	\begin{minipage}{6.5cm}
		\centering
		\includegraphics[scale=.18]{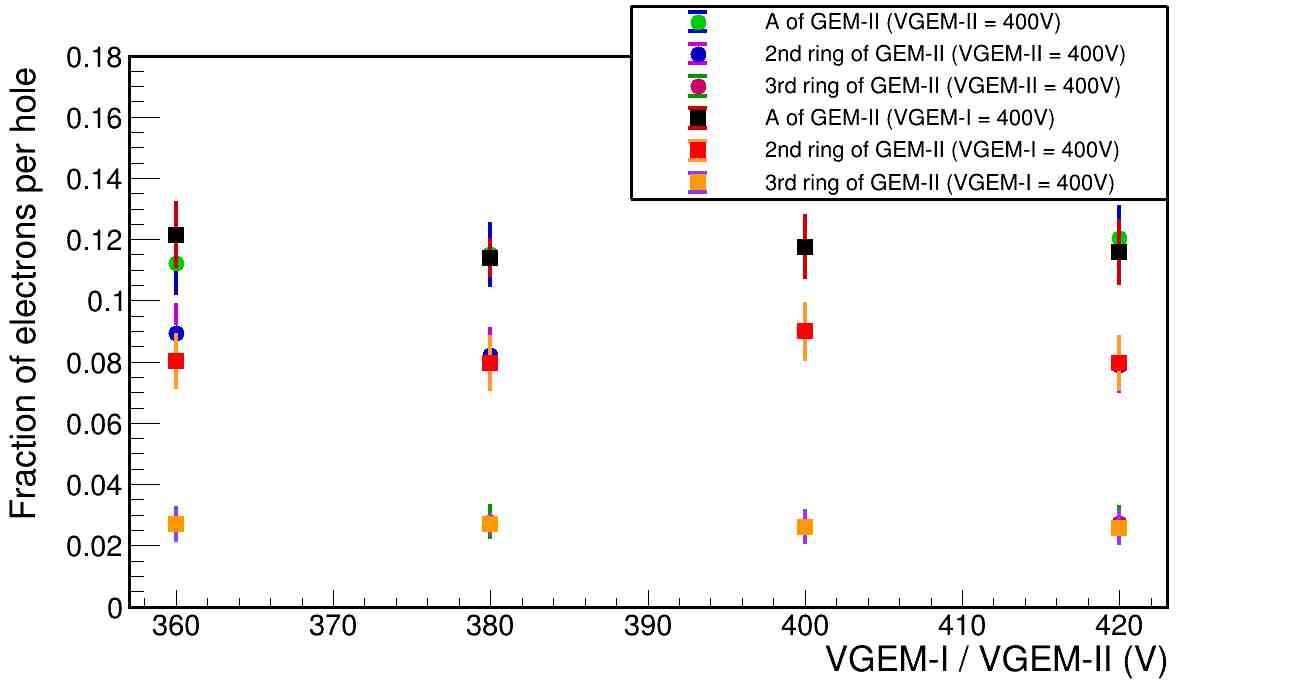}
		\caption{\label{fig7} Fraction of electrons in the central and its adjacent holes in GEM-II for different combinations of VGEM-I and VGEM-II in a double GEM.}
	\end{minipage}
\end{figure}

For a double GEM detector, the top GEM (GEM-I) experiences similar charge distribution as for a single GEM detector. There is one important difference though - the voltage applied across each GEM foil of a double GEM detector is significantly less in comparison to a single GEM detector. As a result, the electron load on the central hole of GEM-I is significantly less than the central hole of a single GEM detector. For instance, it is around 25$\%$ in the central hole and 10$\%$ in the holes of second ring for GEM-I(figure\ref{fig6}). This distribution of charge also depends on the initial position of the electron above holeA. 

For GEM-II of the double GEM detector, the distribution is strikingly different, as shown in figure\ref{fig7}. Although a larger percentage of electrons still enter through the central hole of GEM-II, the fraction is almost half of that in GEM-I and the difference between central and adjacent holes are also reduced in comparison to the earlier situations. For instance, when VGEM-I and VGEM-II both are equal to 400V, the average fraction of electronic charge in the central holeA out of total charge in GEM-II is 11$\%$, in the holes of the second ring (B and others) is 8.5$\%$ and that on the third ring(Q and others) is 2.5$\%$. The number shown in figure\ref{fig7} is the average over 1000 initial electron events. This figure also shows the variation of the fractions when VGEM-I and VGEM-II are varied individually keeping the other one constant. 

It may be noted here, that adding a second GEM foil has led to a more even distribution of electron load among the holes. As a result of this sharing of electron cloud, the maximum effective load per hole is significantly reduced for a double GEM detector, which in turn reduces the probability of discharges.

\begin{figure}[htbp]
	\centering
	\includegraphics[scale=.21]{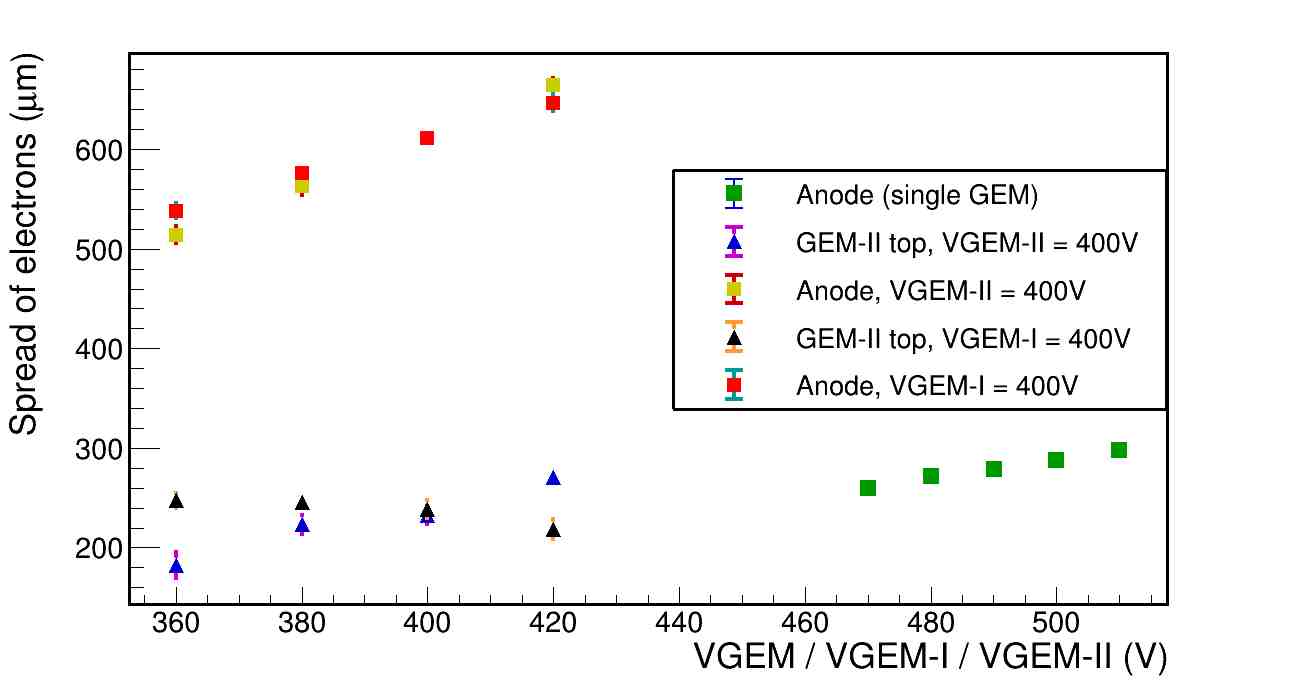}
	\caption{\label{fig8} Spread of electron cloud in single and double GEMs.}
\end{figure}

Spread of electrons at anode increases with increase in GEM voltages for both single and double GEM, as shown in figure\ref{fig8}. Varying GEM-I voltage at constant GEM-II voltages changes the spread of electrons on top of GEM-II whereas there is little change when GEM-II voltage is varied keeping GEM-I constant which can be attributed to the improved focussing of eletric field. Also, it is interesting to note that the magnitude of electron spread on the top of GEM-II is similar to that of anode spread in single GEMs.

\subsection{Spatial resolution}
\begin{figure}[htbp]
	\centering
	\includegraphics[scale=.22]{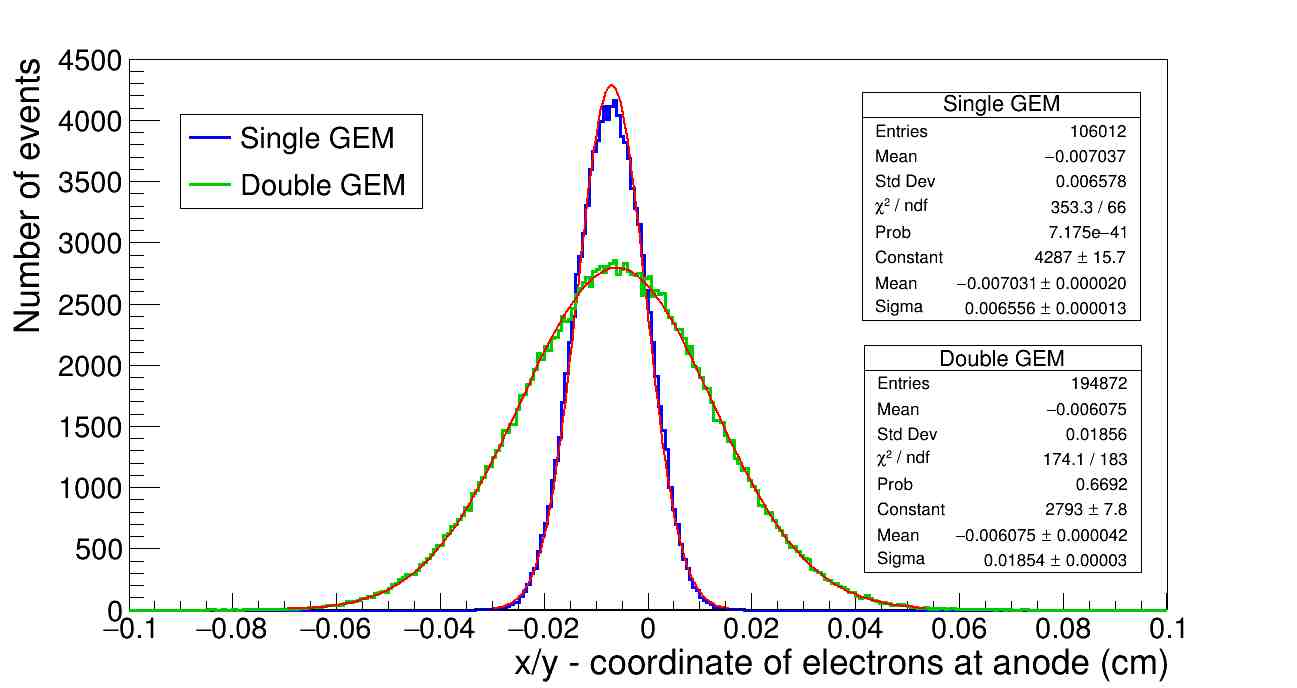}
	\caption{\label{fig9} Distribution of end-points of electrons for single and double GEMs.}
\end{figure}

An intrinsic single point spatial resolution of the detectors under study has been defined to be the sigma of the fitted gaussian distribution of the electron spread at the readout anode. From figure\ref{fig9}, we see that spatial resolution for single GEM is around 65$\mu$m (0.0065 cm) and that of double GEM is 185$\mu$m (0.0185 cm). So, while the sharing of electrons among a larger number of holes in multi-GEM structures reduces the probability of discharges, the position resolution suffers significantly due to the same process.

\section{Conclusion}
Spread of electron cloud at various stages of single and double GEM detectors and their effects have been studied. Addition of multiple stages has been found to influence the spread significantly, and leads to less electron load per GEM hole, in general. Spread of electron cloud at anode is also strongly influenced by the geometry and electric configuration of the detectors. This, in turn, affects the position resolution of the detectors.

\acknowledgments

This work has been partly performed in the framework of the RD51 Collaboration. We wish to acknowledge the members of the RD51 Collaboration for their help and suggestions. We would also like to thank Mr. Jaydeep Datta and Mr. Sridhar Tripathy for presenting the poster in the conference. 



\end{document}